\documentclass[conference]{IEEEtran}
\IEEEoverridecommandlockouts
\usepackage{inputenc}
\usepackage{cite}
\usepackage{amsmath,amssymb,amsfonts}
\usepackage{algorithmic}
\usepackage{graphicx}
\usepackage{textcomp}
\usepackage{xcolor}
\usepackage[normalem]{ulem}
\usepackage{siunitx}
\def\BibTeX{{\rm B\kern-.05em{\sc i\kern-.025em b}\kern-.08em
    T\kern-.1667em\lower.7ex\hbox{E}\kern-.125emX}}
\usepackage{amsmath, amssymb, bm, cite, epsfig, psfrag}
\usepackage{graphicx}
\usepackage[margin=0.7in]{geometry}
\usepackage{dblfloatfix}
\usepackage{array}
\usepackage{lipsum}
\newcolumntype{P}[1]{>{\centering\hspace{0pt}}p{#1}}
\newcolumntype{M}[1]{>{\centering\hspace{0pt}}m{#1}}
\newcolumntype{L}{>{\centering\arraybackslash}m{3cm}}
\usepackage[font=small]{caption}
\usepackage{epstopdf}	
\usepackage{longtable}
\usepackage{supertabular,booktabs}
\usepackage{bbm}
\usepackage{multirow}

\usepackage{subfigure}
\usepackage{etoolbox}
\usepackage{pbox}
\usepackage{tabu}	
\usepackage{filecontents}
\usepackage{enumerate}
\usepackage{textcomp}

\usepackage{colortbl}
\usepackage{fancyhdr}

\usepackage{bm}
\newtoggle{conference}
\togglefalse{conference} 
\makeatletter
\newsavebox\myboxA
\newsavebox\myboxB
\newlength\mylenA
\newcommand*\xoverline[2][0.75]{%
    \sbox{\myboxA}{$\m@th#2$}%
    \setbox\myboxB\null
    \ht\myboxB=\ht\myboxA%
    \dp\myboxB=\dp\myboxA%
    \wd\myboxB=#1\wd\myboxA
    \sbox\myboxB{$\m@th\overline{\copy\myboxB}$}
    \setlength\mylenA{\the\wd\myboxA}
    \addtolength\mylenA{-\the\wd\myboxB}%
    \ifdim\wd\myboxB<\wd\myboxA%
       \rlap{\hskip 0.5\mylenA\usebox\myboxB}{\usebox\myboxA}%
    \else
        \hskip -0.5\mylenA\rlap{\usebox\myboxA}{\hskip 0.5\mylenA\usebox\myboxB}%
    \fi}

\newcommand{\tabincell}[2]{\begin{tabular}{@{}#1@{}}#2\end{tabular}}

\fancypagestyle{firststyle}{
	\fancyhf{}
	\fancyhead[L]{S. Ju, S. Shah, M. Javed, J. Li, G. Palteru, J. Robin, Y. Xing, O. Kanhere, and T. S. Rappaport, ``Scattering Mechanisms and Modeling for Terahertz Wireless Communications,''  2019 \textit{IEEE International Communications Conference (ICC)}, Shanghai, China, May. 2019, pp. 1-7.}     

}

\begin{document}

\title{Scattering Mechanisms and Modeling for Terahertz Wireless Communications}

\author{\IEEEauthorblockN{Shihao Ju, Syed Hashim Ali Shah, Muhammad Affan Javed, Jun Li, Girish Palteru, Jyotish Robin, \\
Yunchou Xing, Ojas Kanhere, Theodore S. Rappaport}
\IEEEauthorblockA{\textit{NYU WIRELESS,}
\textit{NYU Tandon School of Engineering} \\
\{shao, shs515, maj407, jl7333,  gp1492, jr4954, ychou, ojask, tsr\}@nyu.edu}
\thanks{This research is supported by the NYU WIRELESS Industrial Affiliates Program and NSF Grants 1702967 and 1731290.}
}

\maketitle
\thispagestyle{firststyle}
\begin{abstract}
This paper provides an analysis of radio wave scattering for frequencies ranging from the microwave to the Terahertz band (e.g., 1 GHz - 1 THz), by studying the scattering power reradiated from various types of materials with different surface roughnesses. First, fundamentals of scattering and reflection are developed and explained for use in wireless mobile radio, and the effect of scattering on the reflection coefficient for rough surfaces is investigated. Received power is derived using two popular scattering models - the directive scattering (DS) model and the radar cross section (RCS) model through simulations over a wide range of frequencies, materials, and orientations for the two models, and measurements confirm the accuracy of the DS model at 140 GHz. This paper shows that scattering can become a prominent propagation mechanism as frequencies extend to millimeter-wave (mmWave) and beyond, but at other times can be treated like simple reflection. Knowledge of scattering effects is critical for appropriate and realistic channel models, which further support the development of massive multiple input-multiple output (MIMO) techniques, localization, ray tracing tool design, and imaging for future 5G and 6G wireless systems.

\end{abstract}


\section{Introduction}~\label{sec:intro}
There has been a tremendous increase in the demand of wireless bandwidth for a multitude of data heavy applications. Recently, millimeter-wave (mmWave) communications was ratified by international standards such as the Third Generation Partnership Program (3GPP) Release 15 \cite{3GPP.25.104} as a key technology to be deployed in the fifth-generation (5G) cellular systems. Future wireless systems are likely to continue to increase in carrier frequency, approaching the Terahertz frequency range \cite{Hirata06a,Kleine11a}. Terahertz communications can provide unprecedented bandwidth that ranges from tens of GHz to hundreds of GHz \cite{Akyildiz14a}, but relatively little is known in the communications literature about Terahertz frequencies (e.g., from 300 GHz to 3 THz) \cite{rappaport19a}. 

At conventional microwave frequencies (from 300 MHz to 3 GHz), scattering is a much weaker propagation mechanism than reflection and diffraction \cite{Rap02a}. Scattering is typically ignored in current wireless communication system design, and has not been well investigated for wireless communication applications. At mmWave and THz frequencies, the wavelength becomes smaller than the typical surface roughness of many objects, which suggests that illuminated scatterers may give rise to signal paths that are comparable to the power of reflected paths in particular directions \cite{Rap15a}. Thus, a stronger scattering effect can partially compensate for other propagation losses as frequencies increase above 30 GHz. 

Three-dimensional (3D) scattering models will play a key role in the design of 3D ray tracing techniques that will become more important for site-specific mmWave and Terahertz system designs \cite{Haddad11a}. To simulate an accurate radio propagation environment at mmWave and Terahertz frequencies, the power carried by scattered rays needs to be considered, especially in non-line-of-sight (NLOS) conditions where the scattered rays may act as the dominant propagation paths to support a communication link \cite{Rap02a}. 

This paper is organized as follows: Section \ref{early} reviews existing scattering models, including early scattering research used in radar applications. Section \ref{model} compares two popular scattering models,  the DS model and the RCS model, and clarifies the assumptions and possible use cases for each. The numerical results obtained from the models for a wide range of practical materials and surface roughness are discussed and compared with measurements in Section \ref{simulations}. Concluding remarks are given in Section \ref{conclusion}.

\section{Fundamentals of Scattering}~\label{early}
\begin{figure}   
	\centering
	\includegraphics[width=0.8\linewidth]{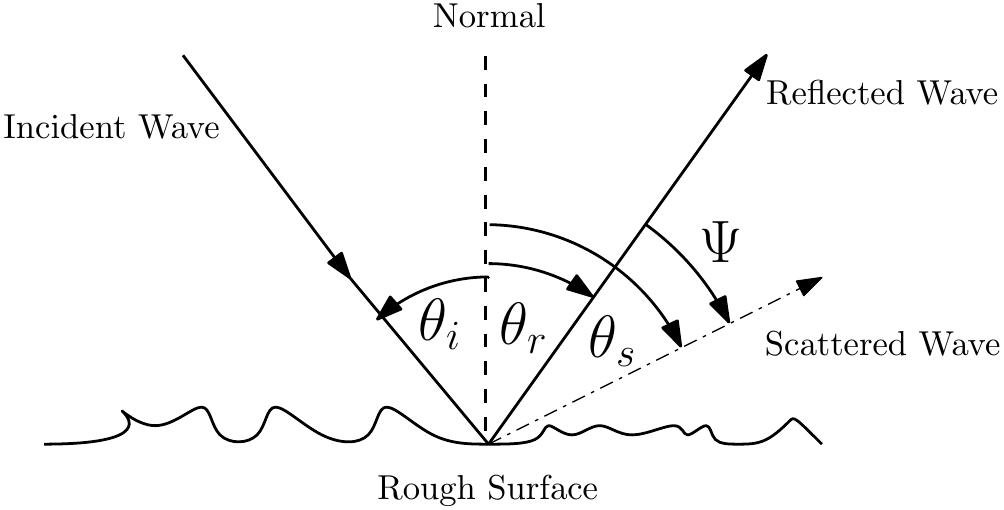}
	\caption{ A radio wave incident at an angle $\theta_{i}$ with respect to the normal to the targeted rough surface. $\theta_{r}$ and $\theta_{s}$ are reflected and scattered angle, respectively. From Snell's law, reflections obey $\theta_{i} = \theta_{r}$. $\Psi$ is the angle between reflected and scattered waves.}
	\label{fig:reflection}
	\vspace{-1.5em}
\end{figure}

The fundamentals of scattering and reflection are unified by geometric optics. Consider the geometry shown in Fig. \ref{fig:reflection} where an electromagnetic wave impinges upon a surface at an incident angle of $\theta_i$. If the surface is smooth and in the far field, and is electrically large compared to the wavelength, no scattering occurs, and reflection may be the only radiation from the surface. On the other hand, if the surface is rough, in addition to a primary reflected component in the specular direction, the incident wave is also scattered into many other directions. This phenomenon is called \textit{diffuse scattering}.

Since the smoothness or roughness of a surface determines the degree of scattering that occurs, it is imperative to ascertain a methodology for defining the smoothness of a surface. Intuitively, the surface roughness should be relative to the wavelength of the impinging wave. At lower frequency range (longer wavelengths), scattering is often negligible as any protuberances on the surface are very small compared to the wavelength of the impinging wave \cite{Rap02a}, and most propagation paths occur due to reflection, where energy from a surface is simply reflected at angle $\theta_r$ in Fig. \ref{fig:reflection}. However, at frequencies of hundreds of GHz, most of the surfaces in a physical environment may be considered rough, since the wavelength becomes comparable to the surface roughness and an impinging wave at angle $\theta_i$ reradiates energy from the rough surface at many angles, not just the reflected angle $\theta_r$ in Fig. \ref{fig:reflection}. Thus, understanding the properties of surfaces is essential in identifying the propagation channel characteristics for future mmWave and Terahertz communication systems \cite{Kleine11a}. 


In order to determine the smoothness or roughness of a surface, the Rayleigh Criterion is used. The Rayleigh Criterion defines the critical height of a surface, $h_c$, which depends on the incident angle and frequency of the incident wave upon the surface. $h_c$ is given by \cite{Rap02a}:

\begin{equation}
    h_c = \frac{\lambda}{8\cos{\theta_i}} 
    \label{eq:hc}
\end{equation}
where $\theta_i$ is the incident angle and $\lambda$ is the wavelength. If the minimum to maximum surface protuberance, $h_0$, is smaller than $h_c$, the surface is considered smooth for a particular $\lambda$ (e.g., frequency); if $h_0$ is larger than $h_c$, the surface is considered rough for a particular $\lambda$ \cite{Rap02a}. 
The surface roughness is usually modeled by two parameters, the root mean squared (rms) height of the surface ($h_{\text{rms}}$) and the correlation length ($l_c$) \cite{Mittal10a}. $h_{\text{rms}}$ is a measure of the vertical roughness of the surface and is given by:
\begin{equation}
h_{\text{rms}}=(\xoverline{h^2}-\xoverline{h}^2)^{(1/2)}
\end{equation}
where $h$ is the surface height relative to the mean surface height ($\bar{h}$). $h$ is usually modeled as a Gaussian random variable with standard deviation $h_{\text{rms}}$ \cite{Jansen11a}. $1/l_c$ is a measure of the horizontal roughness of the surface, where $l_c$ is the correlation length beyond which two points on the surface are considered statistically independent of surface heights \cite{Mittal10a}. 

\subsection{Rough Surface Reflection Coefficient Model}
For a smooth surface, the incident wave reflects off the surface into the specular direction as determined by Snell's law, and there is no scattering. The reflection coefficient of a smooth surface, $\Gamma_{\text{smooth}}$, determines the fraction of incident field that is reflected into the specular direction \cite{Rap02a} (eq. (4.24), pp.116). Note that $\Gamma_{\text{smooth}}$ incorporates any penetration loss that occurs due to part of the electromagnetic wave penetrating the surface \cite{Rap02a}. The power of a wave reflected off a rough surface is diminished due to a portion of the incident power being reradiated in various directions via scattering. Thus, in order to incorporate this \textit{scattering loss} in the reflected wave, a new reflection coefficient is defined for the rough surface, $\Gamma_{\text{rough}}$, to characterize the reflected energy, and other reradiated energy is due to scattering in other directions. $\Gamma_{\text{rough}}$ is obtained by multiplying a scattering loss factor, $\rho_s$, with the reflection coefficient, $\Gamma_{\text{smooth}}$, of the corresponding smooth surface as given by \cite{Rap02a} (eq. (4.24), pp.116):
\begin{equation}
    \Gamma_{\text{rough}}=\rho_s \cdot \Gamma_{\text{smooth}} ,
    \label{eq:gamma}
\end{equation}
where the scattering loss factor, $\rho_s$, is approximated by \cite{Ament53a}:
\begin{equation}
    \begin{split}
        \rho_s=\exp{\Big[-8\left(\frac{\pi h_{\text{rms}} \cos{\theta_i}}{\lambda}\right)^2}\Big],
    \end{split}
    \label{eq:rho_s1}
\end{equation}
where $|\Gamma_{\text{rough}}|\leq 1$, $\rho_s\leq 1$, $h_{\text{rms}}$ is the rms height of the surface, and $\lambda$ is the wavelength. A more precise modeling \cite{Rap02a, Boithias87a} for the scattering loss factor is given as:
\begin{equation}
    \begin{split}
        \rho_s=\exp{\Big[-8\left(\frac{\pi h_{\text{rms}} \cos{\theta_i}}{\lambda}\right)^2}\Big]I_0\Big[8\left( \frac{\pi h_{\text{rms}} \cos{\theta_i}}{\lambda}\right)\Big],
    \end{split}
    \label{eq:rho_s2}
\end{equation}
where $I_0$ is the zero-order Bessel function of the first kind. The rough surface reflection coefficient with scattering loss factor can provide an accurate prediction of the received power in a reflected direction in the design of a ray tracer. It is important to note here that this scattering loss factor, $\rho_s$, is used to determine the \textit{loss} that can be expected in the \textit{reflected direction} due to the scattering phenomenon. It does not give any insight into the strength of the scattered power in a particular direction. 

There are several important things to note about (\ref{eq:gamma})-(\ref{eq:rho_s2}). First, (\ref{eq:rho_s1})-(\ref{eq:rho_s2}) is also regarded as a Rayleigh roughness factor in an extension of Kirchhoff scattering theory called the Beckmann-Kirchhoff theory \cite{Beckmann87a}, and is applicable to both dielectric and metallic surfaces. Second, (\ref{eq:rho_s1})-(\ref{eq:rho_s2}) is based on an assumption that the incident angle $\theta_i$ is small, which means the reflected power due to the scattering loss predicted by (\ref{eq:gamma}) may not be reliable when the incident angle is large (e.g., a small grazing angle). Further, the Beckmann-Kirchhoff scattering theory for rough surfaces is extended to arbitrary incident and scattering angles, not just the reflected angle \cite{Cynthia98a}, where the scattered power predicted by the extended model shows a good agreement with experimental scattering data at many scattering angles when the incident angle $\theta_i$ is large (e.g., $\ang{70}$). 

\section{Scattering Models}~\label{model}
In this paper, two common scattering models are compared. The first model, DS model, is widely used in optics \cite{Navarro17a}. The second model, RCS model, is well known in radar applications. RCS model relies on empirical values based on field measurements. In \cite{seidel1991path}, measurements at 900 MHz in four European cities provided a range of RCS values of common scatterers in typical cellular channels. An empirical approximation of RCS based on the RCS measurements of a number of buildings is $4A$ where $A$ is the projected surface of the building which can be ``seen'' from the receiver (RX) \cite{Rees87a}. We compare DS and RCS models at frequencies of 1, 10, 100, and 1000 GHz, for three exemplar materials with different surface roughness ($h_{rms}$), scattering coefficient ($S$), correlation length ($l_c$), scattering lobe width parameter ($\alpha_R$), and dielectric constant ($\epsilon_r$) as shown in Table \ref{tab:srp}.

\begin{table}
\centering
\caption{\textsc{Exemplar Materials and Parameters}}
\label{tab:srp}
\begin{tabular}{|c|c|c|c|c|c|}
\hline
Materials & $h_{\text{rms}}$ ($\mu$m) & $S$ & $l_c$ ($\mu$m) & $\epsilon_r$ & $\alpha_R$\\
\hline
Material 1 - Smooth & 10 & 0.05 & 1000 & 16 &1\\
\hline
Material 2 - Intermediate	& 100 & 0.3 & 500 & 4&1\\
\hline
Material 3 - Rough& 300 & 0.5 & 300 & 2&1\\
\hline
\end{tabular}
\vspace{-1.5em}
\end{table}

\subsection {Directive Scattering Model}

\begin{figure}[b]   
	\centering
	\includegraphics[width=0.45\textwidth]{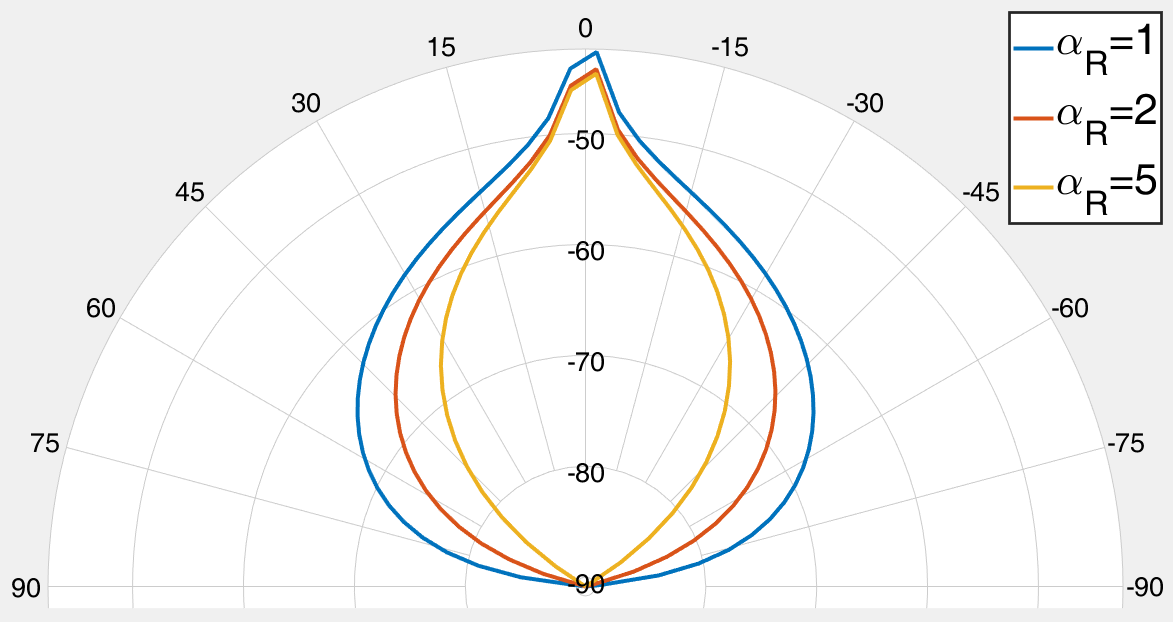}
	\caption{The scattering patterns with different values of $\alpha_R$. $\alpha_R$ determines the width of the scattering lobe. A smaller $\alpha_R$ denotes a wider scattering lobe.}
	\label{fig:alpha_r}
	\vspace{-1.5em}
\end{figure}

The single-lobe DS model \cite{ Esposti07a,Jarvelainen12} assumes that the main scattering lobe is steered in the direction of the specular reflection\textcolor{black}{, and ignores backscatter. The DS dual lobe model includes both forward scatter and backscatter in (\ref{equ:dual}). The single-lobe model is first introduced in this section.} The DS model has been used in \cite{Jarvelainen12} to model the RF propagation environment of a hospital room at 60 GHz. The power delay profile (PDP) of the environment agreed well with simulations using the DS model, up to an excess delay of 30 ns. The DS model has also been tested at 1.296 GHz in \cite{Esposti07a}, where the DS model agreed with the scattering from rural and suburban buildings. When an electromagnetic wave impinges upon a surface at an incident angle $\theta_i$, the scattered electric field at any particular scattering angle $\theta_s$ can be calculated using the DS model. Note that we only consider the incident angle $\theta_i \in [\ang{0},\ang{90}]$. The scattering angle $\theta_s \in [\ang{-90},\ang{90}]$ such that the scattered wave may have any arbitrary direction in the entire incident plane (e.g., including backscattering when $\Psi$ is less than $-\theta_r$ in Fig. \ref{fig:reflection}). The single-lobe DS  scattered electric field in the incident plane is given by:
\begin{align}
\begin{split}
         |\textbf{E}_s|^2 &= |\textbf{E}_{s0}|^2 \cdot \left(\frac{1+\cos(\Psi)}{2} \right)^{\alpha_R}\\
         &=\left(\frac{SK}{d_t d_r} \right)^2 \frac{l\cos{\theta_i}}{F_{\alpha_R}} \cdot \left(\frac{1+\cos(\Psi)}{2} \right)^{\alpha_R}
\end{split}
\label{eq:Es}
\end{align}
where $\textbf{E}_s$ is the scattered electric field at the scattering angle $\Psi$. $\textbf{E}_{s0}$ is the maximum scattered electric field, which is adopted from an effective roughness model\cite{Esposti07a}. S is the \textit{scattering coefficient} \cite{Navarro17a}, which is defined as the fraction of incident electromagnetic wave that is scattered, can be written as $S=|\textbf{E}_s|/|\textbf{E}_i|$, where $|\textbf{E}_i|/|\textbf{E}_0|\propto1/d$, and $d$ is the distance (in meters) between the transmitter (TX) and scattering rough surface. $|\textbf{E}_0|$, $|\textbf{E}_i|$, $|\textbf{E}_s|$ are the magnitudes of transmitted, incident, and scattered electric fields, respectively, in units of V/m. K is given by $\sqrt{60 P_t G_t}$ which is a constant depending on the transmitted power and the TX antenna gain \cite{Esposti01a}. $d_t$ and $d_r$ are the distances between the scatterer and the TX and RX, respectively. $l$ is the length of the scattering object. $\Psi$ is the angle between the reflected wave and the scattered wave, as shown in Fig. \ref{fig:reflection}. $\alpha_R$ is a parameter which determines the width of the scattering lobe, and Fig. \ref{fig:alpha_r} shows that higher values of $\alpha_R$ imply that the scattering lobe is narrower. $F_{\alpha_R}$ is a scaling factor to normalize to the total scattered power, which is given by \cite{Esposti07a}
\begin{align}
\begin{split}
         F_{\alpha_R} = \int_{-\pi/2}^{\pi/2} \left( \frac{1+\cos{\Psi}}{2} \right)^{\alpha_R} \sin{\theta_s}d\theta_s
\end{split}
\label{eq:Far}
\end{align}
where $\theta_s$ is the scattered angle as shown in Fig. \ref{fig:reflection}.

\begin{table}
\centering
\caption{\textsc{Simulated differences between scattered power and reflected power in the specular reflection direction at 500 GHz using (\ref{eq:gamma}-\ref{eq:pr_directive}).}\\\footnotesize{$P_t=$10 W, $A_e = 5$ cm$^2$, $\alpha_R=$1, $l=$10 m, $S=$0.3, and $d_t=d_r=$10 m.}}
\label{tab:sr}
\begin{tabular}{|c|c|c|c|c|}
\hline
$\theta_i$ & Materials&\tabincell{c}{Reflection\\power (dBm)} & \tabincell{c}{Scattered\\ power (dBm)} & \tabincell{c}{Difference\\ (dB)}\\ \hline
\multirow{3}{*}{\ang{0}} & Smooth&  -6.21 &  -53.92& 47.71\\
 & Interm. &  -32.07 & -41.88& 9.81\\
 & Rough & -188.35 & -33.92 &-154.37\\
 \hline
\multirow{3}{*}{\ang{30}} & Smooth&  -5.58 &  -69.12 & 63.54 \\
 & Interm. &  -25.94 & -57.08 & 31.14\\
 & Rough & -143.78 & -49.12 & -94.67\\
\hline
\multirow{3}{*}{\ang{45}} & Smooth & -4.83 & -71.50 & 66.67 \\
 &Interm.& -19.47 &  -59.46 & 39.99\\
 & Rough&  -98.75 & -51.50 & -47.25\\
 \hline
 \multirow{3}{*}{\ang{60}} & Smooth&  -3.87 &  -73.89 & 70.02 \\
 & Interm. &  -12.37 & -61.85 & 49.48\\
 & Rough & -52.81 & -53.89 & 1.08\\
\hline
\multirow{3}{*}{\ang{90}} & Smooth & -1.58 & negligible & Reflection Only \\
 & Interm.&  -1.58 & negligible & Reflection Only\\
 & Rough& -1.58 & negligible & Reflection Only\\ \hline
\end{tabular}
\vspace{-1.5em}
\end{table}
The received power at the RX can then be calculated as \cite{Rap02a}:
\begin{align}
     P_r = P_dA_e = \frac{|\textbf{E}_s|^2}{120\pi}\cdot\frac{G_r\lambda^2}{4\pi} = \frac{|\textbf{E}_s|^2G_r\lambda^2}{480\pi^2}
     \label{eq:pr_directive}
\end{align}
where $P_d$ is the power flux density of the scattered wave, and $A_e$ is RX antenna aperture \cite{Rap02a}. $G_r$ is the RX antenna gain, and $\lambda$ is the wavelength of the radio wave. Table \ref{tab:sr} indicates how the scattered power and the reflected (specular) power change with the incident angle $\theta_i$ when $\Psi=0$ (which indicates that the scattered power is calculated in the direction of the reflected wave) for three materials at 500 GHz. The scattered power is calculated using (\ref{eq:Es})-(\ref{eq:pr_directive}), and the reflected power is calculated using (\ref{eq:gamma}) and (\ref{eq:rho_s2}). When $\theta_i$ = \ang{1}, \ang{30}, and \ang{45} (small incident angles), \textit{the scattered power in the reflected direction is stronger than the reflected power for Material 3 (rough surface)}, as seen in Table \ref{tab:sr}. 
 
\subsection {Radar Cross Section Model}

\begin{figure}   
	\centering
	\includegraphics[width=0.8\linewidth]{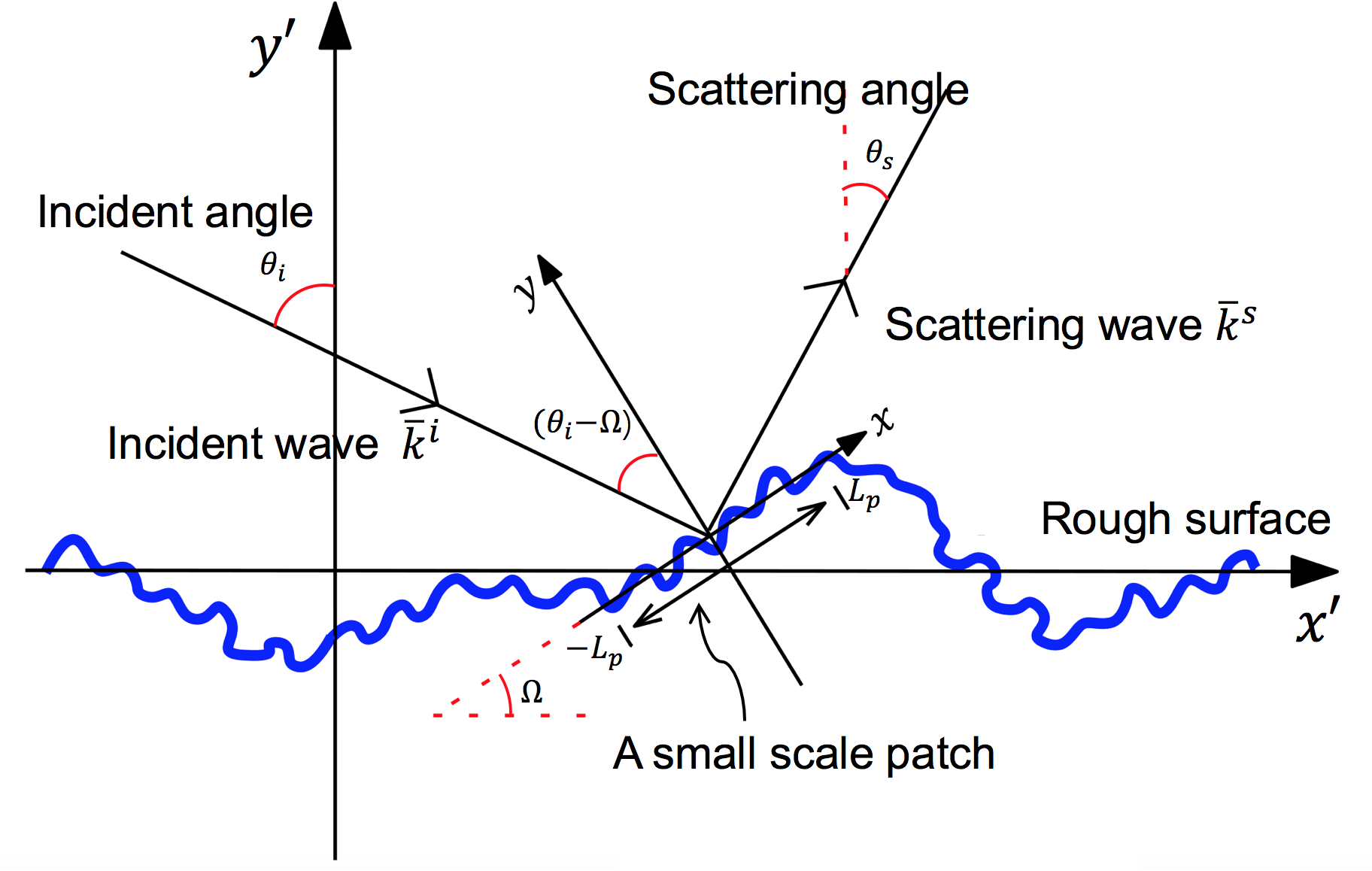}
	\caption{The entire surface is embedded in the $x'y'$ coordinate system. The mean value of height of the rough surface is $y'=0$. Each small patch has its $xy$ coordinate system \cite{Bahar96a}. Each patch has a $x$ range of [$-L_p, L_p$] with slope $\tan\Omega$.}
	\label{fig:RCS}
	\vspace{-1.5em}
\end{figure}

The RCS ($\sigma$) of a scattering object is a measure of the power density scattered in the direction of the RX relative to the power density of the radio wave illuminating the scattering object \cite{Balanis12a}. The RCS model traces its origin to radar theory where it was originally designed to detect large, metallic objects such as aircrafts and ships in the far-field \cite{Skolnik80a}. Thus, the RCS model usually assumes that the scattering object is a perfect electrical conductor, which may not be true for practical wireless environment. Useful insights can be gained about the impact of scattering in a propagation environment by applying the RCS model, where RCS values for different scattering objects are generally empirically derived from measurements \cite{Rap02a}\cite{Rap15a}\cite{seidel1991path}.

Depending on relative positions of the TX and RX, RCS model can be either monostatic or bistatic. The monostatic RCS model describes how the field is scattered in the direction of the RX when the TX and RX are co-located spatially. Thus, computing the received power translates to calculating the reradiated scattered power directly back to the TX (backscattered power) \cite{Skolnik80a}. In contrast, the bistatic RCS model can describe more general cases and measures the electric field scattered in the direction of the RX when the TX and RX may not be co-located. Analysis presented in this paper uses the monostatic RCS model for simplicity \cite{Rap02a}\cite{Rap15a}. 

The received power at the RX that is co-located with the TX is derived using the monostatic RCS model as follows. The directional radiated power density from the TX is:
\begin{align}
    P_d=\frac{P_t G}{4\pi d^2}
\end{align}
where $P_t$ is the transmitted power, $G$ is the gain of the TX antenna, and $d$ is the distance between the TX and scatterer. 

The scatterer intercepts a portion of the incident power and reradiates it in different directions. The measure of the amount of incident power intercepted by the scatterer and reradiated back in the direction of the RX is denoted as the RCS $\sigma$ \cite{Balanis12a}, and the power density at the RX is:
\begin{equation}
    P_{dR} = \frac{P_tG}{4\pi d^2} \cdot \frac{\sigma}{4 \pi d^2} 
    \label{eq:P_dR}
\end{equation}
Given the RX antenna aperture, $A_e$, the received power is:
\begin{equation}
    \begin{split}
    P_r &= P_{dR}\cdot A_e= \frac{P_tG}{4 \pi d^2}\cdot \frac{\sigma}{4\pi d^2} \cdot \frac{G \lambda^2}{4 \pi}= \frac{P_t G^2 \lambda^2 \sigma}{(4\pi)^3 d^4}
    \end{split}
\end{equation}
or, in dB, is given as:
\begin{equation}
    \begin{split}
        P_{R}[\text{dBm}] =& P_{T}[\text{dBm}] + 2G[\text{dBi}] +   20\log(\lambda) + \sigma[\text{dB}\cdot \text{m}^2] \\
        &- 30\log(4\pi) - 40\log(d)
    \end{split}
    \label{eq:pr_rcs}
\end{equation}

RCS has been used to characterize scatterers in a number of empirical studies \cite{Yamada05,Bahar81}. In \cite{seidel1991path}, bistatic RCS measurements of common scatterers were conducted at 900 MHz. The bistatic RCS values were shown to vary between -4.5 and 55.7 dBm$^2$. \textcolor{black}{Monostatic RCS of pedestrians was measured to be -8 dBm$^2$ at 76 GHz, 20 dB lower than the RCS of vehicles \cite{Yamada05}. Composite rough surfaces were modeled with RCS in \cite{Bahar81}.}



The RCS can be thought of as a combination of contributions from two classes of roughness: small scale and large scale. As explained in \cite{JSJTP98}, \textcolor{black}{large-scale and small-scale surface roughnesses are like ocean waves, some waves span hundreds of meters, some waves are very short, less than a meter. Thus, the scattering from the surface can be calculated by dividing the surface into  small-scale and large-scale patches.} A typical small-scale patch is shown in Fig. \ref{fig:RCS}. 

Thus as shown in \cite{Rap15a, BaharShi98} :
\begin{equation}\label{eq:rcs_tot}
    \begin{split}
        \text{RCS}=\sigma=\sigma_{\text{rough}}+|\chi_s|^2 \sigma_{\text{smooth}}
    \end{split}
\end{equation}
where $\sigma_{\text{rough}} $ and $\sigma_{\text{smooth}}$ correspond to the small-scale and large-scale RCS, respectively. In (\ref{eq:P_dR})-(\ref{eq:pr_rcs}), $\sigma$ represents the three-dimensional RCS (dB$\cdot$m$^2$) which denotes an area intercepting the amount of power that, when scattered, produces at the RX a density that is equal to the density reradiated by the scatterer \cite{Balanis12a}. In this paper, we assume that the scattering object has unit length along one dimension for simplicity. Therefore, the three-dimensional RCS can be simplified as the RCS per unit length (two-dimensional RCS) with the unit of dB$\cdot$m. Thus, $\sigma$, $\sigma_\text{smooth}$, and $\sigma_\text{rough}$ with the unit of dB$\cdot$m are used in the rest of this paper.

$\sigma_{\text{smooth}}$ depends on the large-scale dimensions of the scattering object, where only the width of the scattering object is considered since the length of the scattering object is assumed to have unit length. $\sigma_{\text{smooth}}$ is given by \cite{Balanis12a}: 
\begin{equation}\label{eq:sigma_smooth}
    \begin{split}
        \sigma_{\text{smooth}}=\frac{2\pi w^2}{\lambda}\bigg[\cos(\theta_i)\frac{\sin(k_0w\cos(\theta_i))}{k_0w \cos(\theta_i)}\bigg]^2
    \end{split}
\end{equation}
which $w$ is the width of the scattering object, and $k_0=2\pi/\lambda$ is the free space wave number. (\ref{eq:sigma_smooth}) indicates that $\sigma_{\text{smooth}}$ follows an inverse relation with the wavelength of the incident radio wave. 


The weighting factor $\chi_s$ in (\ref{eq:rcs_tot}) is the rough surface height characteristic function, which is given in \cite{Rap15a}:
 \begin{equation}
    \begin{split}
        \chi_s=e^{-k_0^2h_{\text{rms}}^2\cos^2(\theta_i)}
    \end{split}
    \label{eq:chi_ss}
\end{equation}
where  $h_{\text{rms}}$ denotes the rms height of the small-scale surface. The weighting factor $\chi_s$ approaches 1 as frequency decreases, which implies that $\sigma_{\text{smooth}}$ dominates the RCS $\sigma$. However, as frequency reaches the Terahertz band, $\chi_s$ becomes negligible for rough surfaces, and the impact of $\sigma_{\text{rough}}$ on the RCS $\sigma$ becomes much more significant.

$\sigma_{\text{rough}}$ can be obtained by calculating the weighted average cross sections of the individual, randomly orientated small patches as shown in Fig. \ref{fig:RCS}. The orientations of the small scale patches are characterized by $h_x$, $h_x=\tan\Omega$ is shown in Fig. \ref{fig:RCS}. Additionally, $h_x$ is assumed here to be Gaussian \cite{BaharShi98, Bahar93a}:
\begin{equation}
    \begin{split}
        p(h_x)=  \frac{1}{\sqrt{2\pi}\sigma_l}e^{\frac{-h_x^2}{2\sigma_l^2}}
    \end{split}
    \label{eq:pdf_hx}
\end{equation}

The scattering cross section for a small-scale patch with a specific slope $h_x$ is given by \cite{BaharShi98}:
\begin{equation}
    \begin{split}
        \sigma^{PP}_0=|S^{PP}(\bar{k}^s,\bar{k}^i)|^2 Q(\bar{k}^s,\bar{k}^i) 
    \end{split}
    \label{eq:sigma_rough_per_unit}
\end{equation}
where $S^{PP}(\bar{k}^s,\bar{k}^i)$ denotes the scattering coefficient at the patch surface as shown in Fig. \ref{fig:RCS}. $\bar{k}^s$ and $\bar{k}^i$ are the wavevectors of the scattering wave and incident wave, respectively. $PP=HH$ or $VV$ where $H$ denotes horizontal polarization, and $V$ denotes vertical polarization. $Q(\bar{k}^s,\bar{k}^i)$ is obtained from a complete expansion of the fields and the boundary conditions.

Then $\sigma_{\text{rough}}$ is modulated by the slope of small-scale patches by using (\ref{eq:pdf_hx}) and (\ref{eq:sigma_rough_per_unit}):
\begin{equation}
    \begin{split}
        \sigma_{\text{rough}}^{PP}&=\int_{-\infty}^{\infty}\sigma^{PP}_0 p(h_x)dh_x\\ &=\int_{-\infty}^{\infty}|S^{PP}(\bar{k}^s,\bar{k}^i)|^2 Q(\bar{k}^s,\bar{k}^i) p(h_x)dh_x  
    \end{split}
    \label{eq:sigma_rough}
\end{equation}

For a monostatic radar, $\bar{k}^s=-\bar{k}^i$, so the scattering coefficient in (\ref{eq:sigma_rough}) is given as \cite{BaharShi98}:
\begin{equation}
    \begin{split}
        S^{VV}(-\bar{k}^i,\bar{k}^i)&=2(1+\sin^2(\theta_i-\Omega)) \\
        S^{HH}(-\bar{k}^i,\bar{k}^i)&=2\cos^2(\theta_i-\Omega)
    \end{split}
\end{equation}
where $\theta_i$ is the incident angle and $\Omega$ is the tilt angle of the small patch as shown in Fig. \ref{fig:RCS}.

The function $Q(\bar{k}^i,\bar{k}^i)$ is obtained by using the surface height characteristic function $\chi(v_y)$ and joint characteristic function $\chi_2(v_y,-v_y)$ \cite{BaharShi98}.
\begin{equation}
    \begin{split}
        Q(-\bar{k}^i,\bar{k}^i)= &\frac{k_0^3(1+h_x^2)}{v_y^2}\int_{-L_p}^{L_p}(1-\frac{|x|}{L_p})[\chi_2(v_y,-v_y) \\
        &-|\chi(v_y)|^2]e^{-jv_xx}dx
    \end{split}
\end{equation}

Assuming the rough surface heights follow a Gaussian distribution \cite{BaharShi98, Bahar93a}:
\begin{equation}
    \begin{split}
        |\chi(v_y)|^2&=e^{-v_y^2h_{\text{rms}}^2} \\\
        \chi_2(v_y,-v_y)&=|\chi(v_y)|^2e^{v_y^2h_{\text{rms}}^2R_s(x)}
    \end{split}
    \label{eq:chi2}
\end{equation}
where 
\begin{equation}
        h_{\text{rms}}^2R_s(x)=h_{\text{rms}}^2e^{-\frac{x^2}{l_c^2}}\\
    \label{eq:chi23}
\end{equation}
As shown in Fig. \ref{fig:RCS}, $\bar{v}=\bar{k}^s-\bar{k}^i=-\bar{k}^i-\bar{k}^i=-2\bar{k}^i$ and it is shown that $v_y=2k_0\cos(\theta_i-\Omega)$, $v_x=2k_0\sin(\theta_i-\Omega)$ \cite{Bahar96a}. $l_c$ denotes the correlation length of the surface height. 

To compare to the DS model with the monostatic RCS model, a special case of the DS model is considered in the next section in which the TX and RX are co-located, and $\Psi=-2\theta_i$ as shown in Fig. \ref{fig:reflection}. In the next section, the TX and RX are assumed to have equal and constant antenna effective aperture over frequencies, meaning that the antenna gain will increase as the frequencies increases.

\section{Numerical Results}~\label{simulations}
\vspace{-1.5em}
\begin{figure*}[htbp]
 	\centering
 	\setlength{\abovecaptionskip}{-0.1cm}
 	\setlength{\belowcaptionskip}{-0.5cm}
	\subfigure[Scattered power in incident direction (backscatter) for Material 2 using DS model (\ref{eq:Es})-(\ref{eq:pr_directive}) vs. incident angle $\theta_i$ for four frequencies.]{
		\begin{minipage}[b]{0.42\textwidth}
		\includegraphics[width=1\linewidth]{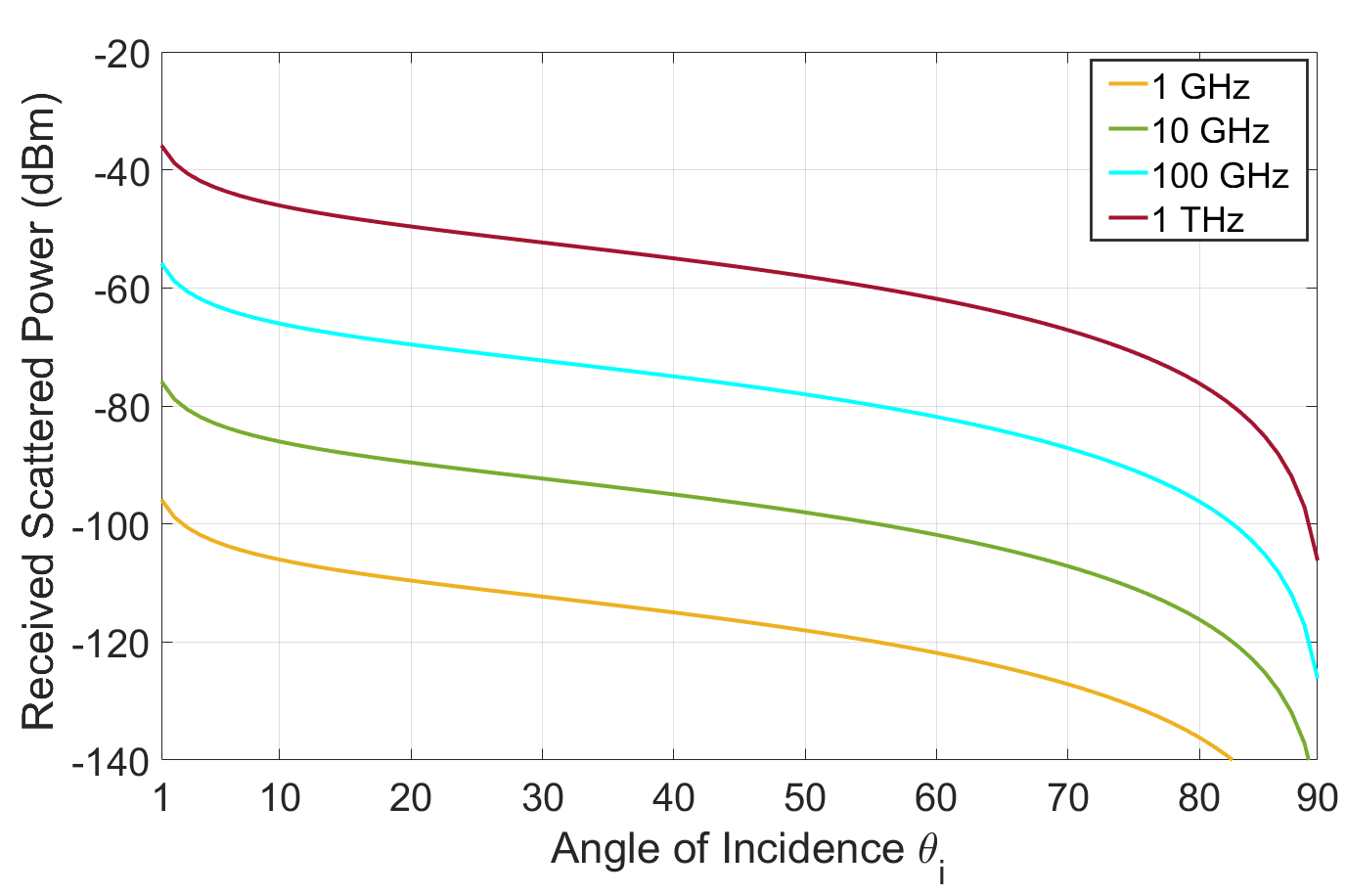}
		\label{fig:ds_freq}
		\vspace{-1.5em}
		\end{minipage}
	}
	\quad
	\subfigure[Scattered power in incident direction (backscatter) for Material 2 using RCS model (\ref{eq:pr_rcs})-(\ref{eq:chi2}) vs. incident angle $\theta_i$ for four frequencies.]{
		\begin{minipage}[b]{0.42\textwidth}
		\includegraphics[width=1\linewidth]{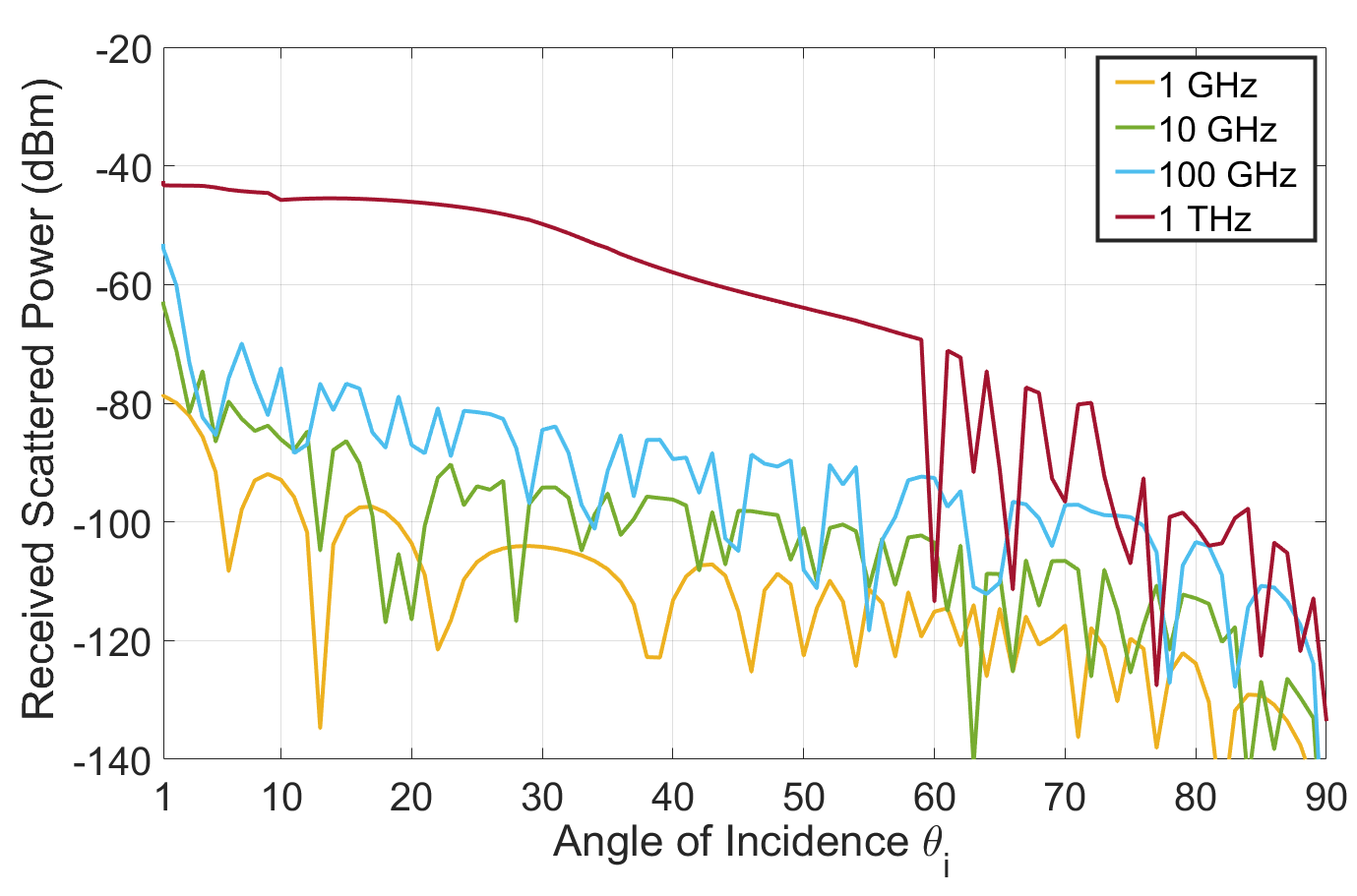}
		\label{fig:rcs_freq}
		\vspace{-1.5em}
		\end{minipage}
	}
	\quad
	\subfigure[Scattered power in incident direction (backscatter) at 100 GHz using DS model (\ref{eq:Es})-(\ref{eq:pr_directive}) vs. incident angle $\theta_i$ for three materials.]{
		\begin{minipage}[b]{0.42\textwidth}
		\includegraphics[width=1\linewidth]{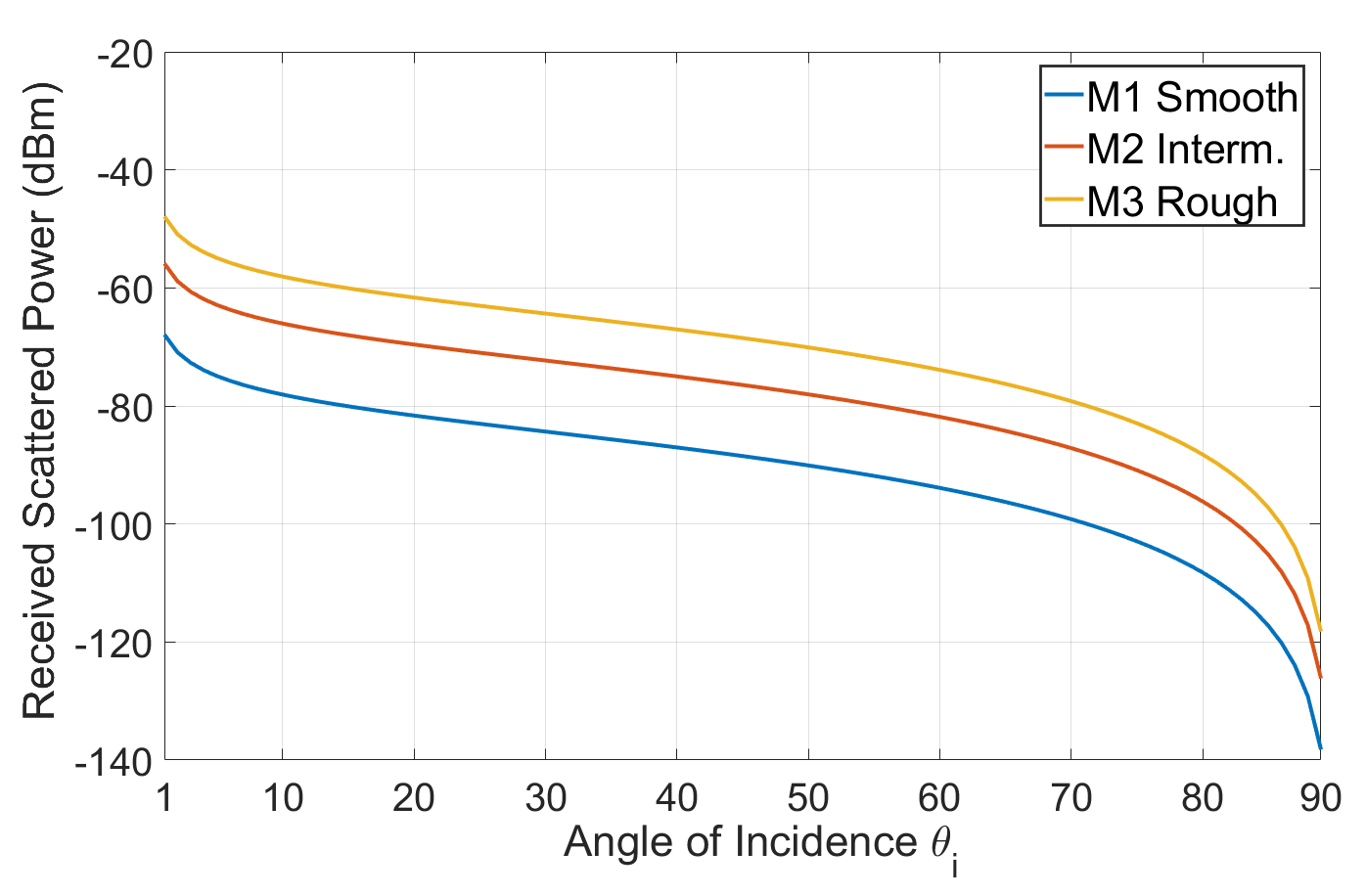}	
		\label{fig:ds_mat}
		\vspace{-1.5em}
		\end{minipage}
	}
	\quad
	\subfigure[Scattered power in incident direction (backscatter) at 100 GHz using RCS model (\ref{eq:pr_rcs})-(\ref{eq:chi2}) vs. incident angle $\theta_i$ for three materials.]{
		\begin{minipage}[b]{0.42\textwidth}
		\includegraphics[width=1\linewidth]{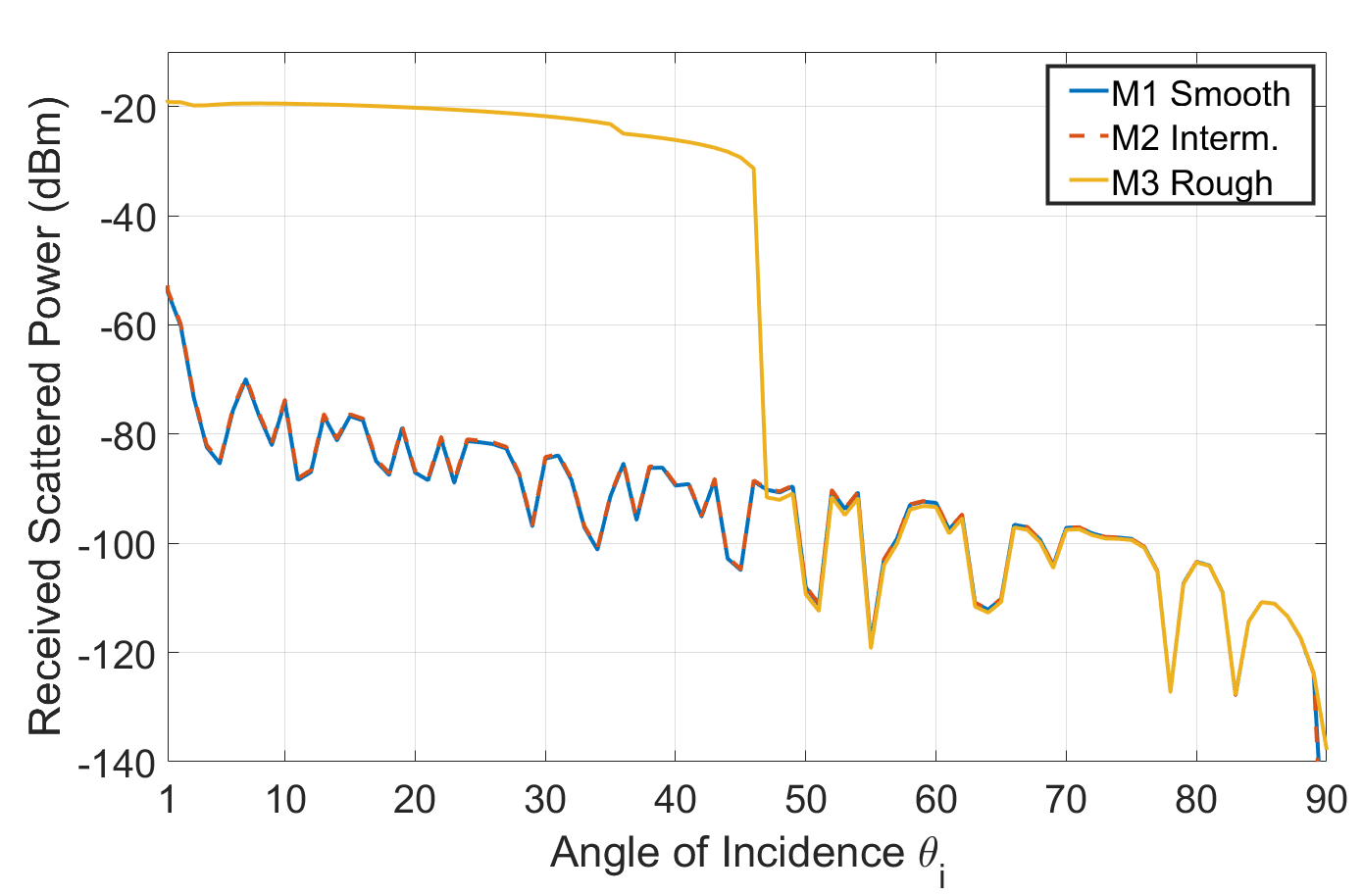}
		\label{fig:rcs_mat}
		\vspace{-1.5em}
		\end{minipage}
	}
	\caption{Scattered power radiated in incident direction (backscatter) using different frequencies and different materials.} 
	\label{ds_and_rcs}
\end{figure*}

Simulations are performed using three materials over frequencies ranging from 1 GHz to 1 THz and incident angles $\theta_i$ ranging from \ang{0} to \ang{90}. (\ref{eq:Es})-(\ref{eq:pr_directive}) are used to calculate the received scattered power from the DS model, and (\ref{eq:pr_rcs})-(\ref{eq:chi2}) are used to calculate the received scattered power from the RCS model. The values of surface roughness applied in simulations are shown in Table. \ref{tab:srp}. The TX and RX are co-located (monostatic) and vertically polarized. The simulations assume a transmit power of 10 W, the antenna apertures of the TX and RX are both kept constant as 5 cm$^2$ over frequencies, TX and RX are 50 m from the scatterer, and the size of scatterer is 1 m x 10 m. 

Fig. \ref{fig:ds_freq} and Fig. \ref{fig:rcs_freq} show how the received scattered power for the monostatic case is dependent on $\theta_{i}$ for both the DS model and the RCS model. Material 2 (Intermediate roughness) is used in the simulations. The backscattered power for the DS model ranges from -30 dBm to -150 dBm, and the received scattered power for the monostatic RCS model ranges from -40 dBm to -150 dBm. As the frequency increases, the received scattered power increases, which can be expected because surfaces tend to be rough as the critical height $h_c$ (in (\ref{eq:hc})) becomes smaller. Moreover, the maximum scattered power is received when the incident wave impinges upon the surface along the normal direction. The scattered power falls off sharply when the incident wave becomes grazing, and most of the incident power is reflected. 

Fig. \ref{fig:ds_mat} and Fig. \ref{fig:rcs_mat} illustrate the variation of the received scattered power for the monostatic case as a function of the incident angle $\theta_{i}$ for three materials at 100 GHz using the DS and the RCS model. Simulation results show that rougher surfaces (higher $h_{\text{rms}}$) cause greater received scattered power. As shown before in Table \ref{tab:sr}, the rough surface can emit more scattered power than reflected power. In Fig. \ref{fig:rcs_mat}, the envelope of the received scattered power follows a decaying profile as the incident angle increases. The received scattered power of the rough surface is much greater than the received scattered power of the other two surfaces at small incident angles because $|\chi_s|$ is small, and $\sigma_{\text{rough}}$ dominates in (\ref{eq:chi_ss}). When the incident angle approaches \ang{90}, Fig. \ref{fig:rcs_mat} shows that the received scattered powers for three different materials start to overlap because $\sigma_{\text{rough}}$ decreases nearly to 0, and thus $\sigma_{\text{smooth}}$ (which is identical for the three materials as shown in (\ref{eq:sigma_smooth})) dominates the received scattered power. 

\begin{figure}   
	\centering
	\includegraphics[width=0.42\textwidth]{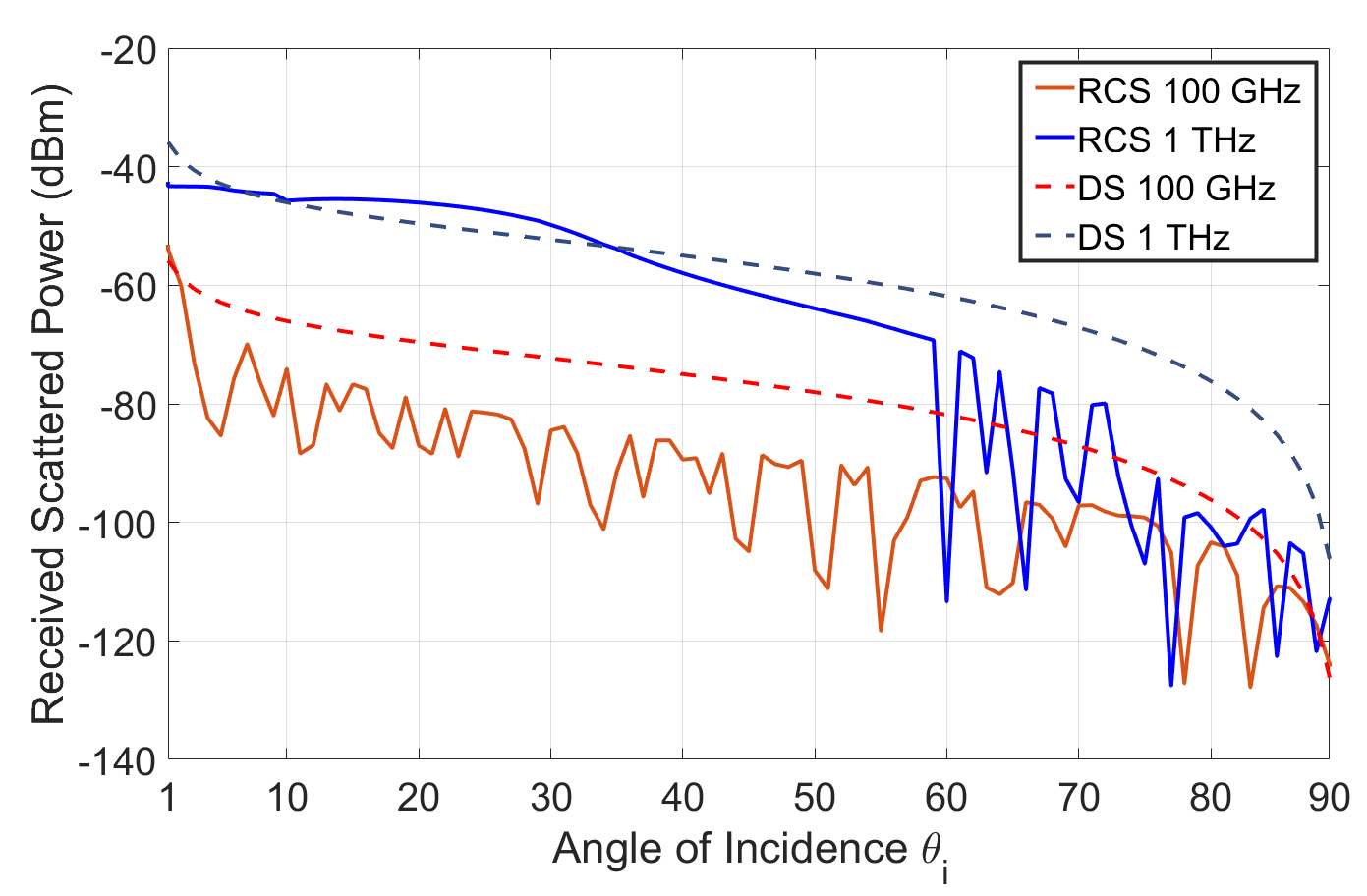}
	\caption{\textcolor{black}{Comparison of backscattered power with $\theta_{i}$ for DS (\ref{eq:Es})-(\ref{eq:pr_directive}) and RCS (\ref{eq:pr_rcs})-(\ref{eq:chi2}) models for Material 2 at 100 GHz and 1 THz.}}
	\label{fig:ds_rcs}
	\vspace{-2em}
\end{figure}

Fig. \ref{fig:ds_rcs} compares the backscattered power for Material 2 at 100 GHz and 1 THz using the DS model and the RCS model. The DS model agrees with the envelope of the RCS model at both 100 GHz and 1 THz as the incident angle $\theta_i$ increases. Since the RCS model is an empirical-based model, the RCS model can be better parameterized to predict the accurate scattered power based on the field measurements for various materials having different surface roughness and permittivity at different frequencies.


To validate the DS model and simulation results, scattering measurements of drywall were conducted at 142 GHz using the channel sounder system introduced in \cite{xing18GC}. During the scattering measurements, both the TX and RX heights were set at 1.2 m on an arc with a radius of 1.5 m using TX output power of -2.35 dBm using 27 dBi gain and 8\textdegree~ half power beam width horn antennas at each end of the link. Incident angles $\theta_i =$ 10\textdegree, 30\textdegree, 60\textdegree, and 80\textdegree~were chosen, and the scattered power was measured from -80\textdegree~to 80\textdegree~in 10\textdegree~steps ($\pm$90\textdegree~could not be measured due to the physical size of the channel sounder system).

\begin{figure}[htbp]
	\centering
 	\includegraphics[width=0.45 \textwidth]{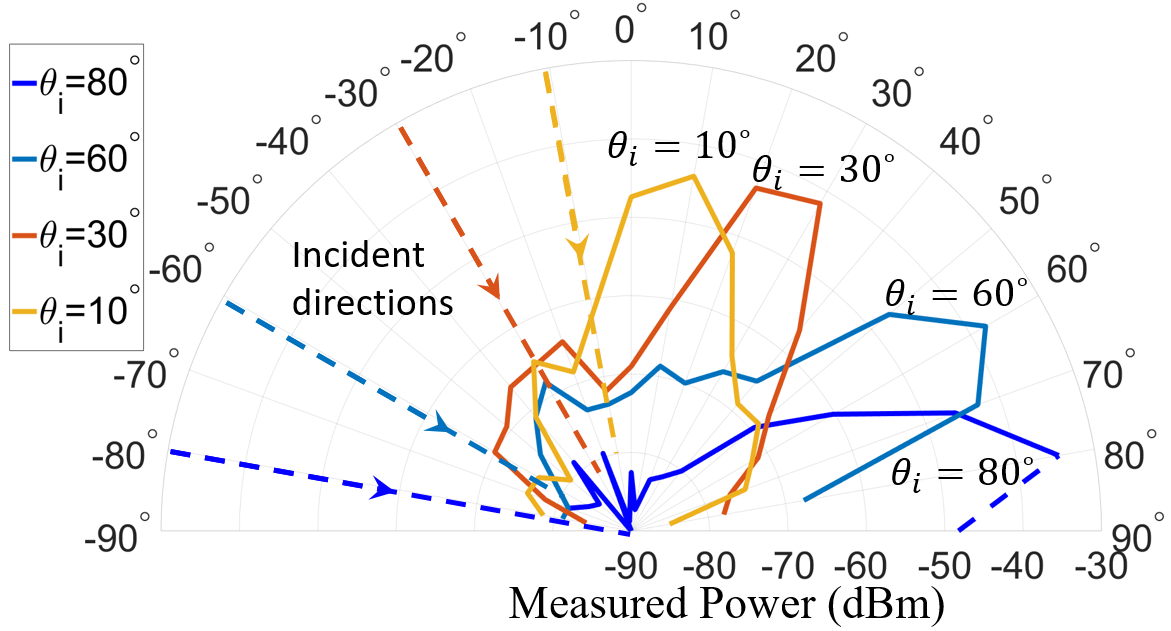}
	\caption{\textcolor{black}{Measured scattered and reflected (total) power from -80\textdegree~to 80\textdegree~from drywall at four incident angles.}}
	\label{fig:scatter_drywall}
	\vspace{-1.5em}
\end{figure}

Measured scattering patterns of different incident angles at 142 GHz are shown in Fig. \ref{fig:scatter_drywall}. The peak measured power (scattered plus reflected) was observed to occuur at the specular reflected angle (Snell's law), and peak measured power was greater at larger incident angles than at smaller incident angles (9.4 dB difference between 80\textdegree~and 10\textdegree), where most of the energy is due to reflection and not scattering. At all angles of incidence, measured power was within 10 dB below the peak power in a $ \pm$ 10\textdegree~angle range of the specular reflection angle, likely a function of antenna patterns. In addition, backscattered power was observed (e.g., 10\textdegree~and 30\textdegree~incidence) but was more than 20 dB below the peak received power.

The single-lobe DS model introduced in Section \ref{model} can be extended to a dual-lobe DS model, which incorporates power of an additional back-scatter lobe \cite{Esposti07a}. The dual-lobe DS scattered electric field is given by:
\vspace{-0.5em}
\begin{eqnarray}~\label{equ:dual}
 |E_s|^2=|\textbf{E}_{s0}|^2 \Bigg[\Lambda\cdot \left(\dfrac{1+\cos(\Psi)}{2} \right)^{\alpha_R}\nonumber\\
 + (1-\Lambda)\cdot \left(\dfrac{1+\cos(\Psi_i)}{2} \right)^{\alpha_i}\Bigg] 
\end{eqnarray}
where $\Psi_i$ is the angle between the scatter ray and the incident ray and $\Lambda$ determines the relative strength of the back-scatter lobe with respect to the forward-scatter lobe.

\begin{figure}[htbp]
 	\centering
  	\setlength{\abovecaptionskip}{-0cm}
 	\setlength{\belowcaptionskip}{-0.5cm}
	\includegraphics[width=0.42 \textwidth]{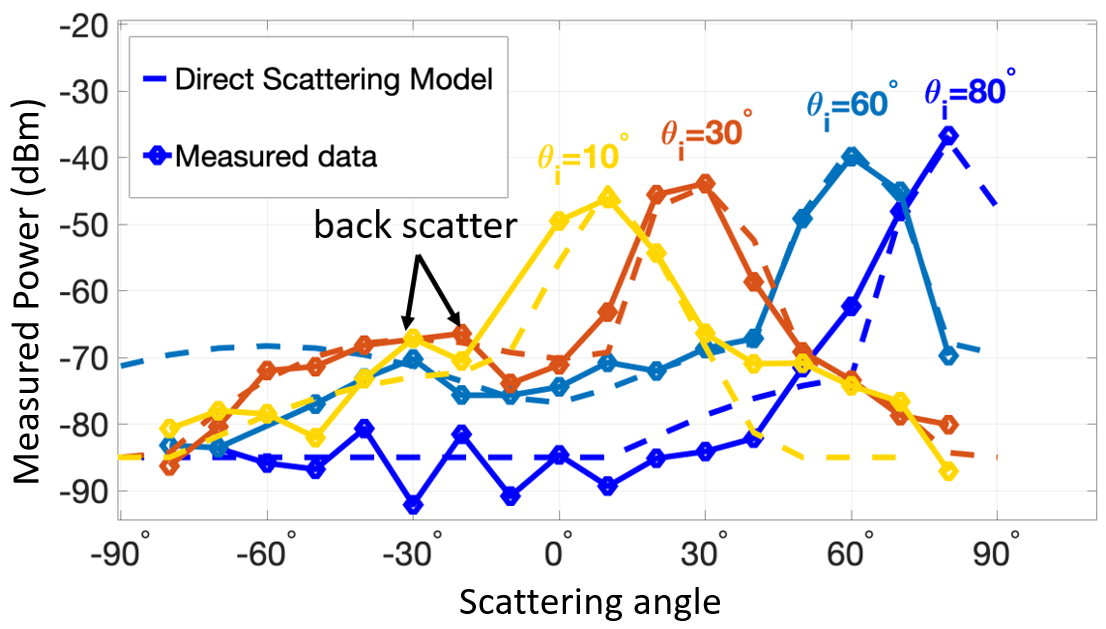}
	\caption{Comparison between measurements and dual-lobe DS model plus reflected power using (\ref{eq:gamma})-(\ref{eq:rho_s2}), (\ref{equ:dual}) at incident angles 10\textdegree, 30\textdegree, 60\textdegree, and 80\textdegree~at 142 GHz for drywall ($\epsilon_r=3.0$ for drywall).}
	\label{fig:TX1080}
\end{figure}

A comparison between measurements and predictions made by the dual-lobe DS model \eqref{equ:dual} with TX incident angle $\theta_i$= 10\textdegree, 30\textdegree, 60\textdegree, and 80\textdegree~is shown in Fig. \ref{fig:TX1080}. The curve is fit for $\Lambda$ ranging from 0.5 (at $\theta_i$ = 30\textdegree) to 0.8 (at $\theta_i$ = 10\textdegree) \cite{Esposti07a}, and simulations of peak received power (the sum of reflection and scattering) at the specular reflection angle agrees well with measured data (within 2 dB), confirming that scattering can be modeled approximately by a smooth reflector with some loss (\ref{eq:gamma})-(\ref{eq:rho_s2}), (\ref{equ:dual}) when material properties are known, while scattering at other angles falls off rapidly. 
\vspace{-0.2cm}
\section{Conclusion}~\label{conclusion}
    This paper investigates scattering using two well-known scattering models, DS and RCS. Simulations of received scattered power for different materials at a wide range of frequencies from 1 GHz to 1 THz with incident angles from \ang{0} to \ang{90} have been conducted. Simulations show that the backscattered power (monostatic case) predicted by the two models decreases as the incident angle becomes larger (close to grazing). The scattered power relative to the reflected power increases with frequency and surface roughness. Notably, the forward scattered power in the reflected direction for a rough surface close to normal incidence (e.g., $\theta_i =$ 0\textdegree) is predicted, via simulations, to be stronger than the specular reflected power, and when close to grazing, scattered power decreases as reflected power dominates \cite{Rap02a,Rap15a}. Thus, scattering at Terahertz depends on surfaces and impinging angles. The peak measured powers (scattered plus reflected) in the reflected direction are greater at larger incident angles, and the peak predicted powers (using DS model plus the rough surface reflected power) are within 2 dB of the peak measured powers at 142 GHz for drywall. Further, backscatter is both modeled and measured to be more than 20 dB down from the peak received power (scattered plus reflected) at small incident angles (e.g., 30\textdegree~or less). To a first order approximation, the models and measurements at 142 GHz show that smooth surfaces like drywall may be modeled as reflective surfaces, especially close to grazing.

\bibliographystyle{IEEEtran}
\bibliography{THz_scatter}
\end{document}